\documentclass[12pt,a4paper]{article}
\usepackage{graphicx}
\usepackage{dcolumn}
\usepackage{bm}
\usepackage{amsmath,amsfonts,amssymb}
\usepackage{slashed}
\usepackage{braket,xcolor}
\usepackage{verbatim}
\usepackage{subcaption}
\usepackage{multirow}
\usepackage{amsfonts,mathtools,resizegather}
\usepackage[utf8]{inputenc}
\usepackage{caption,jheppub}

\title{\boldmath {Bath deformations, islands and holographic complexity
} }
\author[a]{Aranya Bhattacharya,}
\author[b]{Arpan Bhattacharyya,}
\author[a]{Pratik Nandy,}
\author[c]{Ayan K. Patra}

\affiliation[a]{Centre for High Energy Physics, Indian Institute of Science,\\ C.V. Raman Avenue, Bangalore 560012, India.}
\affiliation[b]{\textit{Indian Institute of Technology, Gandhinagar, Gujarat-382355, India}}
\affiliation[c]{Theory Division, Saha Institute of Nuclear Physics, HBNI, 1/AF Bidhannagar, Kolkata 700064, India}
\emailAdd{aranyab@iisc.ac.in}
\emailAdd{abhattacharyya@iitgn.ac.in}
\emailAdd{pratiknandy@iisc.ac.in}
\emailAdd{ayan.patra@saha.ac.in}

\abstract{Considering a doubly holographic model, we study the evolution of holographic subregion complexity corresponding to deformations of bath state by a relevant scalar operator, which corresponds to a renormalization group flow from the AdS-Schwarzschild to the Kasner universe in the bulk. The subregion complexity shows a discontinuous jump at Page time at a fixed perturbation, where the discontinuity depends solely on the system's parameters. We show that the amount of discontinuity decreases with the perturbation as well as with the scaling dimension of the relevant scalar operator.}

\begin{document}
\maketitle
\flushbottom
\section{Introduction}
The black hole information loss paradox has been a longstanding unsolved problem in theoretical physics for the last few decades. Thanks to the recent progress made in \cite{Penington:2019npb,  Almheiri:2019psf, Almheiri:2019hni, Penington:2019kki, Almheiri:2019qdq,  Almheiri:2019psy, Almheiri:2019yqk} couple of years ago, we now understand a few things better than earlier\footnote{See \cite{Almheiri:2020cfm, Raju:2020smc} for recent reviews.}. In these works, as well as many other ones (see \cite{Hashimoto:2020cas, Anegawa:2020ezn, Alishahiha:2020qza, Gautason:2020tmk, Hartman:2020swn, Hollowood:2020cou, Geng:2020fxl, Geng:2020qvw, Li:2020ceg, Chandrasekaran:2020qtn, Akers:2019nfi, Balasubramanian:2020hfs, Hartman:2020khs, Balasubramanian:2020coy, Akal:2020twv, Akal:2021foz, Kawabata:2021hac, Kawabata:2021vyo, Bhattacharya:2020uun, Bhattacharya:2021jrn, Bhattacharya:2021dnd, Sato:2021ftf, Ling:2020laa, Chen:2020jvn, Chen:2019uhq, Chen:2020uac, Chen:2020hmv, Hollowood:2021nlo, Ghosh:2021axl, Chu:2021gdb, Caceres:2021fuw, Geng:2021hlu, Ahn:2021chg, Krishnan:2020oun, Hollowood:2021wkw, Li:2021dmf, Chen:2019iro, Liu:2020gnp, Akal:2020ujg, Iizuka:2021tut, Krishnan:2020fer, Krishnan:2021faa, Bhattacharya:2020ymw, Basak:2020aaa, KumarBasak:2021rrx, Manu:2020tty, Geng:2021wcq, Geng:2021iyq, Anderson:2020vwi, Caceres:2020jcn, Reyes:2021npy, Neuenfeld:2021bsb, Matsuo:2020ypv, Miyaji:2021lcq, Miyata:2021qsm, Balasubramanian:2021xcm} and references therein), the central idea that has been used is the computation of the entanglement entropy of Hawking radiation using the quantum extremal surface (QES) prescription. The QES prescription is a result of continuous modifications on the primarily known Ryu-Takayanagi (RT) \cite{Ryu:2006bv} prescription of computing entanglement entropy in the gravity side via holography. In case of QES, one computes the extremal surface by using the max-min prescription given in \cite{Engelhardt:2014gca}. In this case, the minimization is done on a generalized entropy functional and the QES prescription suggests \cite{Engelhardt:2014gca, Penington:2019npb, Almheiri:2019hni, Almheiri:2019psf, Almheiri:2020cfm}
	\begin{align} \label{eq1}
	S_{\mathrm{EE}}(\mathcal{R})=\mathrm{min} \bigg\{{\underset{\mathrm{Is}}{\mathrm{ext}}} \Big( S_{\rm{QFT}}(\mathcal{R} \cup \mathrm{Is})+\frac{\mathcal{A}(\partial \, \mathrm{(Is)})}{4\, G_{N}}\Big) \bigg\}.
	\end{align}
Here $\mathcal{R}$ denotes the radiation region and ``Is" is the island region. They are some isolated regions in the bulk, that we will discuss later in detail. The entanglement entropy in LHS is the fine-grained entropy of radiation, whereas the entropy (first term) in RHS is the coarse-grained entropy computed semi-classically. The second term is just the RT prescription to compute the area of the island surface.

The crucial change that QES computations in evaporating and eternal black hole models bring in is to yield a unitarity compatible Page curve for the entanglement entropy of the radiation system. While applying QES to the radiation subsystem, it is useful to consider doubly holographic models \cite{Almheiri:2019hni}. In this scenario, the non-gravitational radiation subsystem is supposed to contain a holographic gravity dual. It is also worth noting that in most doubly holographic models that reproduce unitarity compatible Page curves, the radiation bath region is non-gravitating, and the models carry massive gravitons. However, it is not yet fully understood the situation in the case of a gravitating bath. The tension in this line of debate lies in the idea of Hilbert space factorization in gravity \cite{Raju:2021lwh}.


The islands are the essential features of all these models. These are the bulk regions, completely disconnected from the bulk dual of the radiation, which provides a way of encoding nontrivial black hole degrees of freedom by the radiation subsystem starting from a timescale typically known as the Page time (this is the point in time axis where the growth of entanglement entropy curve stops and goes through a phase transition). In the evolution of entanglement entropy, the Page transition is understood by a change of direction of the entanglement entropy curve. However, the entropy changes continuously in the fine-grained curve.

In a set of recent papers, another interesting quantum information-theoretic quantity, known as the subregion complexity \cite{Alishahiha:2015rta}, has been studied in similar doubly holographic models. However, complexity has been found to capture the Page transition through a discontinuous jump. The subregion complexity signifies the hardness in an optimal construction of the evolving mixed states of the black hole and radiation subsystems. In holographic models, the maximal co-dimension one volume below the Hubeny-Rangamani-Takayanagi (HRT) surface \cite{Hubeny:2007xt} is conjectured to represent the measure of the subregion complexity of the corresponding state. Therefore, the evolution of complexity is different from the entanglement entropy ones and thus can provide a complementary way to understand the Page transition. Hence, it appears to be an interesting observation since it represents the crossover through which new degrees of freedom are shared between the two (sub)systems.


Although the discontinuous nature was found in several papers \cite{Bhattacharya:2020uun, Bhattacharya:2021dnd, Bhattacharya:2021jrn, Sato:2021ftf}\footnote{On general grounds, the discontinuous jump in the subregion complexity in various geometries was observed in \cite{Ben-Ami:2016qex}.}, it was only investigated as a case-by-case basis through numerical studies. Hence, there was no particular way to attribute the amount of discontinuity to the physical parameters of the of system (for example, radiation subsystem size, Page time, etc.).  In this paper, we study the braneworld scenario with one brane in the bulk and considering non-gravitating bath. However, the bulk region is taken differently from that of simple AdS black holes. We take the model studied in \cite{Caceres:2021fuw}, where the bath in the conformal boundary is deformed by a scalar perturbation, resulting in a bulk scalar field. In the absence of a brane, such deformation results in a change of the near singular geometry to that of the so-called Kasner universe.\footnote{Kasner geometry has a long history. See \cite{PhysRevLett.22.1071,  doi:10.1080/00018737000101171, Das:2006dz, Engelhardt:2013jda, Barbon:2015ria, Brandenberger:2016egn,  Bolognesi:2018ion, Hartnoll:2020fhc, Caputa:2021pad} and the references therein.} These are understood as Kasner flows in the holographic renormalization group approach. Such a flow from a UV fixed point on the conformal boundary is induced by a scalar deformation   leading to an IR flow near the horizon which finally accounts for a trans-IR flow towards the near-singularity Kasner universe. Changing the scalar amounts to different coarse-graining of the UV state, resulting in different late time linear growth of the Hartman-Maldacena (HM) surface \cite{Hartman:2013qma} in the interior. This induces a scale $r_{\mathrm{RG}}$, probed by the HM surface. UV physics dominates up to this scale starting from the boundary, whereas trans-IR physics takes over afterward until the singularity region. 
 
 \begin{figure}
    \centering
    \includegraphics[width=0.55\textwidth]{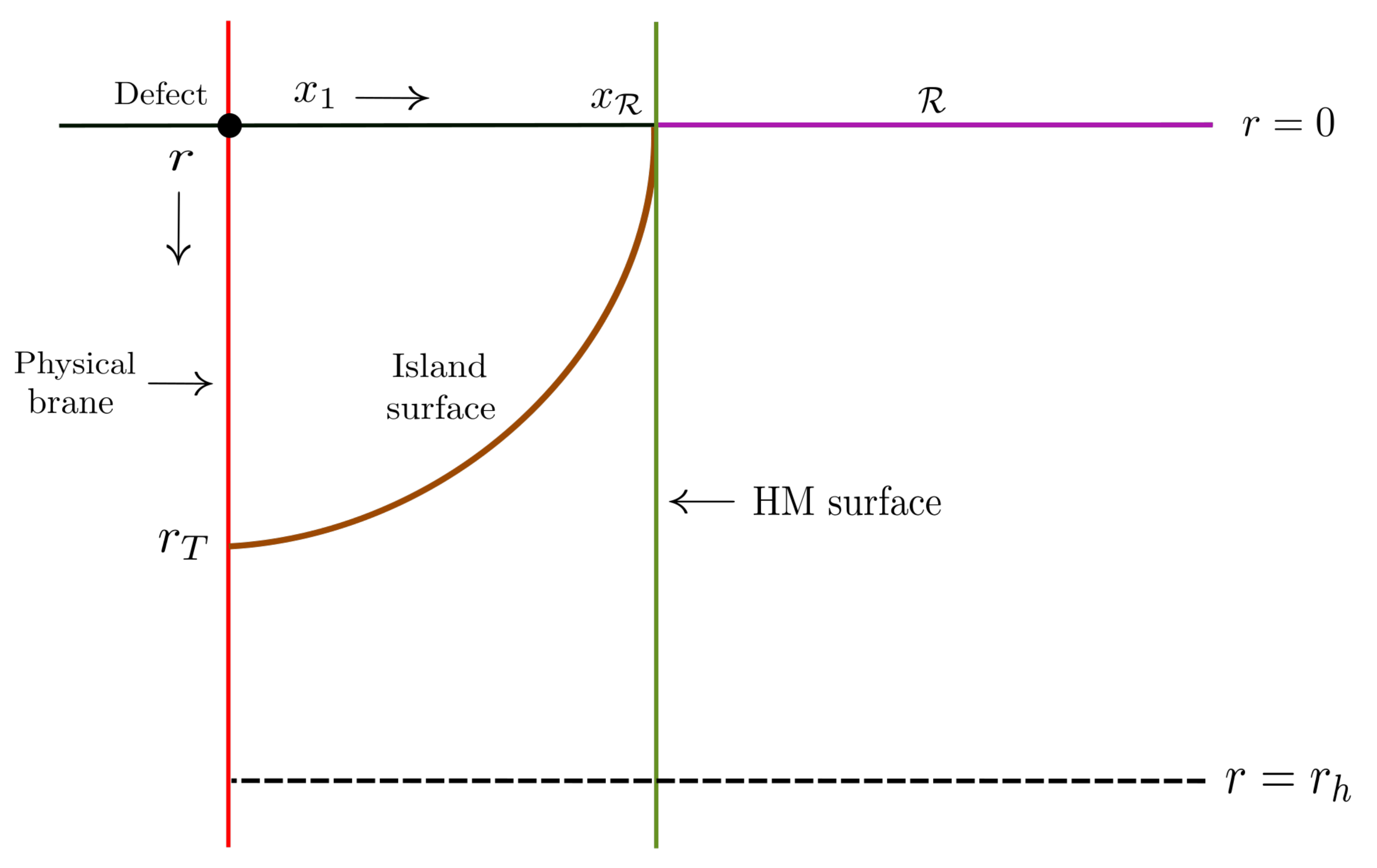}
    \caption{The braneworld model. The left red line is the brane that meets the conformal boundary (black + purple line) at the defect shown in the back dot. The purple line on $r=0$ is the radiation region starting from $x_{\mathrm{R}}$. The green line is the ever-growing Hartman-Maldacena (HM) surface which dominates before the Page time. The brown surface (hits the brane at point $r_T$) is the constant island surface which becomes the minimal surface starting from the Page time.}
    \label{braneworld}
\end{figure}

Finally, upon the introduction of the brane as in the braneworld model under consideration (see Fig.\ref{braneworld}), each of these flows becomes a separate boundary conformal field theory (BCFT) \cite{Takayanagi:2011zk, Fujita:2011fp}\footnote{See \cite{Geng:2021iyq, Miao:2021ual, Miyaji:2021ktr, Hollowood:2021wkw, Chu:2021mvq} for the recent explorations of BCFT.} thermal state parametrized by the deformation parameters. However, since we are interested in the Page curve and therefore need to study the island surfaces as well, another scale $r_{T}$ appears in the picture, which denotes the depth at which the island surface intersects the brane. This scale is determined both by the Kasner exponents and the size of the radiation region considered on the conformal boundary. Crucially, as mentioned in \cite{Caceres:2021fuw}, it is also expected to capture how many degrees of freedom are traced out when one considers an island surface. This interests us as in the subregion complexity studies; it is argued that the jump in complexity at Page transition point is due to purification of degrees of freedom between the radiation degrees of freedom before Page time and new degrees of freedom encoded due to the inclusion of islands. With this motivation and the original one of studying the subregion complexity along a set of renormalization group (RG) flows, we study the volumes under the minimal surfaces. Our motivation is to see if we can find some systematic behavior of the jump at Page time for different scalar perturbations and parametrize the discontinuity in terms of Page time. In this paper, we solely focus on the doubly holographic models which have the following three equivalent renditions \cite{Geng:2020fxl, Caceres:2021fuw}:

\begin{itemize}
\item[(I)]{a BCFT$_d$ i.e., $d$-dimensional BCFT.}
\item[(II)]{a CFT$_d$ coupled to an asymptotically AdS$_d$ gravity, which is further connected to a CFT$_d$. This CFT$_d$ lies on the half line and joined with a transparent boundary conditions to the AdS$_d$ + CFT$_d$.}
\item[(III)]{Einstein gravity which lies on an asymptotically AdS$_{d+1}$ and contains an end-of-the-world (EoW) brane.}
\end{itemize}

The rest of the paper is structured as follows. In section \ref{Kasner}, we review the notion of Kasner universe and RG flow in more detail alongside the minimal area computations. In section \ref{area}, we provide the computations of the HM and island area, and especially the dependence of Page time on the relevant perturbation. Section \ref{complexitybath} contains the detailed calculation of holographic subregion complexity at different times and the corresponding plots showing how the complexity depends on the scalar field deformation. Finally, in section \ref{discussion}, we give a summary of our results and discuss some open questions.

\section{From AdS-Schwarzschild to Kasner universe}\label{Kasner}
In this section, we introduce and review the salient features of the braneworld model, especially the RG flow and the role of Kasner exponents. We closely follow \cite{Frenkel:2020ysx, Caceres:2021fuw, Liu:2021hap}. We set $c = \hbar = 1$ throughout our discussion.

\subsection{Setup}

We take the $(d+1)$-dimensional Einstein-Hilbert action with a negative cosmological constant. We further couple a scalar field with a potential $V(\varphi) = m^2 \varphi^2$ to it\footnote{Adding a self-interacting $\varphi^4$ term has been studied in \cite{Wang:2020nkd}.}. We write the action as \cite{Caceres:2021fuw, Frenkel:2020ysx}
\begin{align}
    I = \int \mathrm{d}^{d+1} x \sqrt{|g|} \left(R + d(d-1) - \frac{1}{2} \nabla^i \varphi \nabla_i \varphi - \frac{1}{2} m^2 \varphi^2 \right).
\end{align}
where we have chosen the normalization $16 \pi G_{d+1} = 1$. We have also taken the cosmological constant as $\Lambda = - d(d-1)/2$ and set $\ell_{\mathrm{AdS}} = 1$. Varying the above action one obtains the equations of motion as \cite{Caceres:2021fuw}
\begin{align}
    G_{i j} - \frac{d(d-1)}{2} g_{i j} &= \frac{1}{4} \Big[2 \, \nabla_i \varphi \nabla_j \varphi - g_{i j} \left(\nabla^i \varphi \nabla_i \varphi  + m^2 \varphi^2 \right) \Big], \label{ein} \\
    (\nabla^i \nabla_i - m^2) \varphi &= 0 \label{KG}.
\end{align}
Here the first equation is obtained by varying the metric which is just the Einstein equation with a scalar coupled to the theory as given by the RHS of Eq.\eqref{ein}. The second equation Eq.\eqref{KG} is the Klein-Gordon equation of the field $\varphi$. We consider the form of the metric \cite{Caceres:2021fuw}
\begin{align}
    \mathrm{d} s^2 = \frac{1}{r^2} \left[ - f(r) e^{-\chi(r)} \mathrm{d} t^2 + \frac{\mathrm{d} r^2}{f(r)} + \mathrm{d} \Vec{x}_{d-1}^2 \right], \label{met}
\end{align}
where $\Vec{x}_{d-1} \in \mathbb{R}^{d-1}$ and $f(r)$ is such that the horizon corresponds to $f(r_{h}) =0$. Our ansatz is that the scalar field $\varphi$ is only dependent on the radial coordinate, i.e., $\varphi = \varphi(r)$. The mass of this operator is related to the scaling dimension of the dual boundary operator via \cite{Aharony:1999ti}
\begin{align}
    \Delta = \frac{d}{2} + \frac{1}{2} \sqrt{d^2 + 4 m^2}, ~~~~\Rightarrow ~~~~ m^2 = \Delta (\Delta -d). \label{mass0}
\end{align}
For the particular case of Schwarzschild solution, we have $\chi (r) = \varphi(r) = 0$ which implies $f(r) = 1 - (r/r_{h})^{d}$. On the other hand, $\varphi(r) = 0$ implies that is there is no back-reaction from the scalar field. This is justified because the Schwarzschild solution is the solution of the vacuum Einstein equation. Further, in our case, it should be noted that the AdS boundary is at $r=0$, while the IR-singularity corresponds to $r \rightarrow \infty$. In some cases, we analytically translate the solution to the trans-IR region to understand the corresponding \emph{Kasner flows}.

With the metric \eqref{met}, the set of differential equations obtained from the equations of motion \eqref{KG} are given by \cite{Caceres:2021fuw}
\begin{align}
    \varphi'' + \left( \frac{f'}{f} - \frac{d-1}{r} - \frac{\chi'}{2} \right) \varphi' - \frac{m^2}{r^2 f} \varphi &= 0, \label{eq1}\\
    \chi' - \frac{2 f'}{f} + \frac{m^2 \varphi^2}{(d-1) r f} - \frac{2 d}{r f} + \frac{2 d}{r} &=0, \label{eq2}\\
    \chi' - \frac{r}{d-1} (\varphi')^2 &=0. \label{eq3}
\end{align}

\subsection{Near boundary and near singularity limits: Kasner exponents}

We can study the near-UV boundary ($r \rightarrow 0$) and near-IR singularity ($r \rightarrow \infty)$ behavior from the above expressions. In particular, the near-singularity behavior leads to the form of the fields
\begin{align}
    \varphi(r) \sim (d-1) \, c \ln r, ~~ \chi(r) \sim (d-1) \, c^2 \ln r + \chi_1,
    ~~ f(r) \sim -f_1 r^{\rho},
\end{align}
where $c$, $\chi_1$ and $f_1$ are constants, and $\rho = d + c^2 (d-1)/2$. The constant $c=0$ leads to the Schwarzschild solution. In this limit the metric \eqref{met} can be recast to the Kasner universe metric of the form
\begin{align}
    \mathrm{d} s^2 = - \mathrm{d} \tau^2 + \tau^{2 p_t} \, \mathrm{d} t^2 + \tau^{2 p_x} \, \mathrm{d} \Vec{x}_{d-1}^2, ~~~~~ \varphi(r) \sim - \sqrt{2}\, p_{\varphi} \ln \tau,
\end{align}
where we have rescaled with the coordinate $r=\tau^{-2/\rho}$ and $p_t, p_{x}$ and $p_{\varphi}$ are known as \emph{Kasner exponents}. They are restricted to satisfy the following set of constraints \cite{Caceres:2021fuw} 
\begin{align}
    p_t + (d-1) p_x = 1,~~~~~ 
    p_{\varphi}^2 + p_t^2 + (d-1) p_x^2 = 1.
\end{align}
The three exponents are constrained by two equations. Hence only one Kasner exponent is free, which is often taken as $p_t$. For our case, we can easily verify the Kasner exponents
\begin{align}
    p_t = 1 - \frac{2 (d-1)}{\rho}, ~~~~~ p_x = \frac{2}{\rho}, ~~~~~ p_{\varphi} = \frac{2 \sqrt{(d-1)(\rho-d)}}{\rho},
\end{align}
satisfy the following Kasner constraints. However, In the Schwarzschild case, $c=0$ implies $\rho =d$, which simplifies the exponents as
\begin{align}
    p_t = - 1 + \frac{2}{\rho}, ~~~~~ p_x = \frac{2}{d}, ~~~~~ p_{\varphi} = 0.
\end{align}
The variation of Kasner exponent $p_t$ with the perturbation is obtained in \cite{Frenkel:2020ysx, Caceres:2021fuw}. In the limit of infinite perturbation, the Kasner exponents are expected to correspond to their respective Schwarzschild values. This also suggests that the Kasner exponents dominate the near-singularity behavior.

The near-boundary behavior of fields is more subtle. First, we note that the mass term should satisfy the Breitenlohner-Freedman bound \cite{Caceres:2021fuw}
\begin{align}
    - \frac{d^2}{4} \leq m^2 < 0. \label{BF}
\end{align}
This bounds gives two alternatives of $\Delta$ by Eq.\eqref{mass0}. However, we will restrict $\Delta$ such that it respects the unitary bound
\begin{align}
    \Delta \geq \frac{d-2}{2}. \label{un}
\end{align}
With this, one can now obtain the near-boundary behavior of $f(r), \varphi(r)$ and $\chi(r)$. However, the behavior of $\varphi(r)$ and $\chi(r)$ will be different for $\Delta = d/2$ and $\Delta \neq d/2$. We list them below \cite{Caceres:2021fuw}
\begin{align}
    f(r) &= e^{\chi(r)} (1 - \braket{T_{tt}} r^d), \\
    \varphi(r) &= 
\begin{cases}
    \varphi_0 r^{d-\Delta}+ \frac{\braket{\mathcal{O}}}{2 \Delta - d} r^d & ~~~~~~~~~~~~~~~~~~~~~~~~~~~~~~~~~~~~~~~~~~~~~\text{if } \Delta \neq d/2\\
    \varphi_0 r^{d/2} \ln r              &~~~~~~~~~~~~~~~~~~~~~~~~~~~~~~~~~~~~~~~~~~~~~ \text{if } \Delta = d/2,
\end{cases} \\
    \chi(r) &= 
\begin{cases}
    \frac{d- \Delta}{2 (d-1)}\varphi_0^2 r^{2(d-\Delta)}+ \frac{2 \Delta (d - \Delta) \braket{\mathcal{O}}}{d (d-1) (2 \Delta - d)} \varphi_0 r^d +  \frac{\Delta \braket{\mathcal{O}}^2}{2 (d-1) (2 \Delta - d)^2} r^{2 \Delta} & \text{if } \Delta \neq d/2\\
    \frac{1}{4 d (d-1)} \varphi_0^2 r^d \, [2 + 2 d \ln r + (d\, \ln r)^2 ]            & \text{if } \Delta = d/2.
\end{cases}
\end{align}
where $\varphi_0$ is the boundary source, $\braket{\mathcal{O}}$ is the one-point function of the boundary operator $\mathcal{O}$, and the energy density of the thermal state is denoted by $\braket{T_{tt}}$. For more details, the readers are referred to \cite{Caceres:2021fuw}.

Our tuning parameter is the dimensionless ratio $\varphi_0/T^{d-\Delta}$, where the temperature is defined through the expression \cite{Caceres:2021fuw}
\begin{align}
    T = \frac{|f_{h}'| e^{-\chi_{h}/2}}{4 \pi}.
\end{align}
Here, $r_{h}$ is the horizon such that $f (r_{h}) = 0$ and $f_{h}' = f (r_{h})$. We have also denoted the abbreviation $\chi_{h}= \chi (r_{h})$. Given that the radial functions are regular at horizon, the near-boundary data of $\braket{T_{tt}}$ and $\braket{\mathcal{O}}$ are equivalent to the ratio $\varphi_0/T^{d-\Delta}$, and this labels the holographic RG flow in the bulk.

With the above boundary data, one solves the equations \eqref{eq1}-\eqref{eq3}. However, the analytically solution is not possible and we need to resort to the numerical methods. We employ the numerical shooting method as discussed in \cite{Caceres:2021fuw}. First, we expand the above functions near horizon limit (i.e., $r \rightarrow r_{h}$ limit) as
\begin{align}
    \varphi(r) &= \varphi_{+} + \varphi'_{+} (r - r_{h}) + O[(r - r_{h})^2], \label{ex1}\\
    f(r) &= f'_{h} (r - r_{h}) + O[(r - r_{h})^2], \label{ex2}\\
    \chi(r) &= \chi_{h} + \chi'_{h} (r - r_{h}) + O[(r - r_{h})^2]. \label{ex3}
\end{align}
where we have used $f(r_{h}) = 0$. Plugging Eq.\eqref{ex1}-\eqref{ex3} in to the Eq.\eqref{eq1}-\eqref{eq3} (multiplying by $r f$ to avoid singularity), we get the following expressions
\begin{align}
    r_{h} f'_{h} \varphi'_{h} + \frac{\Delta (d- \Delta)}{r_{h}} \varphi_{h} &= 0, \\
    -\frac{\Delta (d- \Delta)}{d-1}\varphi_{h}^2 + 2 (d + r_{h} f'_{h}) &= 0, \\
    \chi'_{h} - \frac{r_{h}}{d-1} (\varphi'_{h})^2 &= 0.
\end{align}
The solutions are
\begin{align}
    \varphi_{h} &= \mp \, \frac{i \sqrt{2} \sqrt{d-1} \sqrt{3 + r_{h} f'_{h}}}{\sqrt{\Delta(d-\Delta)}}, \label{s1}\\
    \varphi'_{h} &= \pm \frac{i 2\sqrt{2} \sqrt{d-1} \sqrt{d + r_{h} f'_{h}} \sqrt{\Delta(d-\Delta)} }{r^2_{h} f'_{h}}, \label{s2}\\
    \chi'_{h} &= - \frac{2 [\Delta(d-\Delta)] (d + r_{h} f'_{h})}{r^3_{h} f'^2_{h}}. \label{s3}
\end{align}

 	\begin{figure}
		\centering
		\begin{subfigure}[b]{0.48\textwidth}
			\centering
			\includegraphics[width=\textwidth]{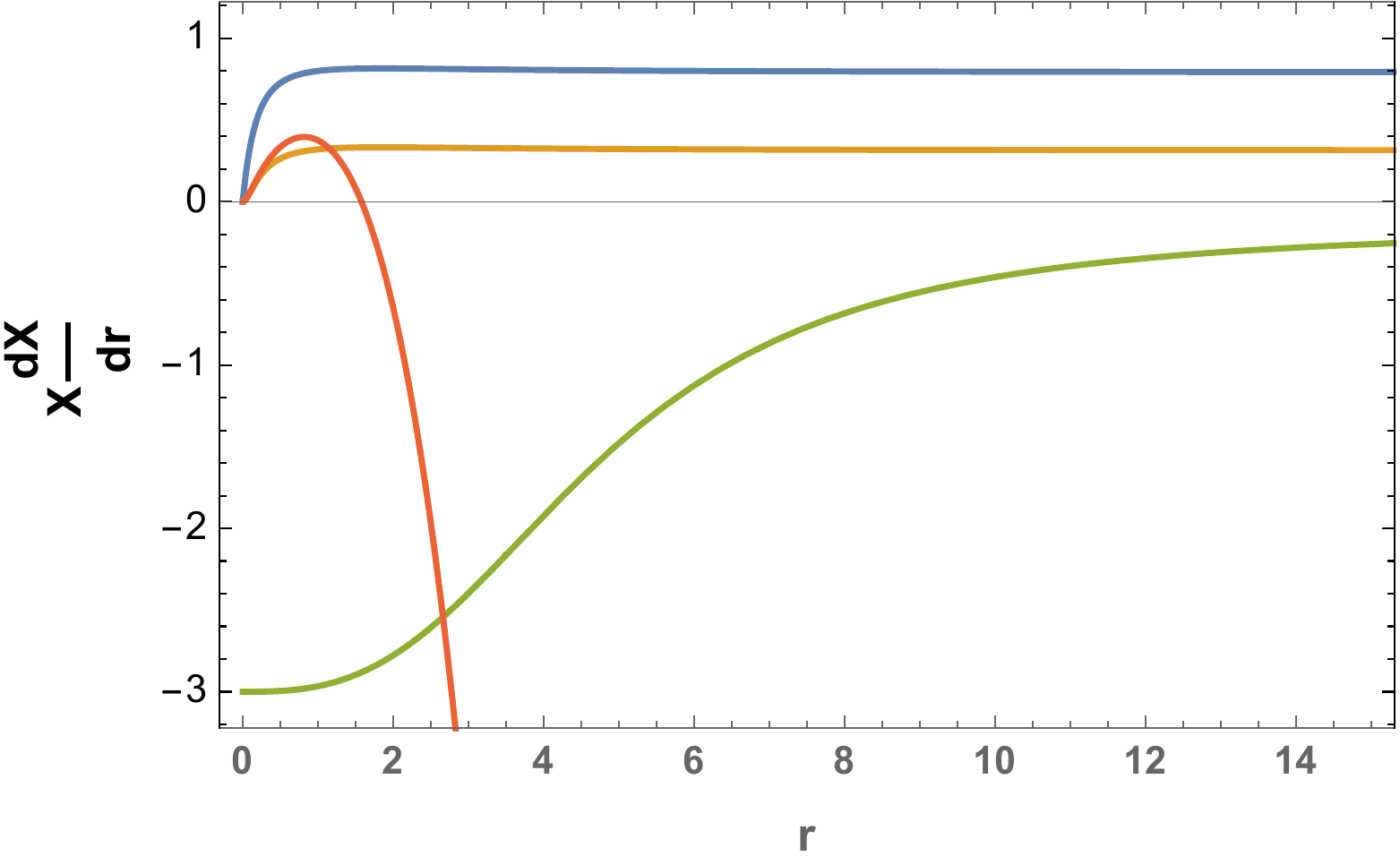}
			\caption{RG flow from AdS to Kasner universe.}
			\label{fig:fl}
		\end{subfigure}
		\hfill
		\begin{subfigure}[b]{0.45\textwidth}
			\centering
			\includegraphics[width=\textwidth]{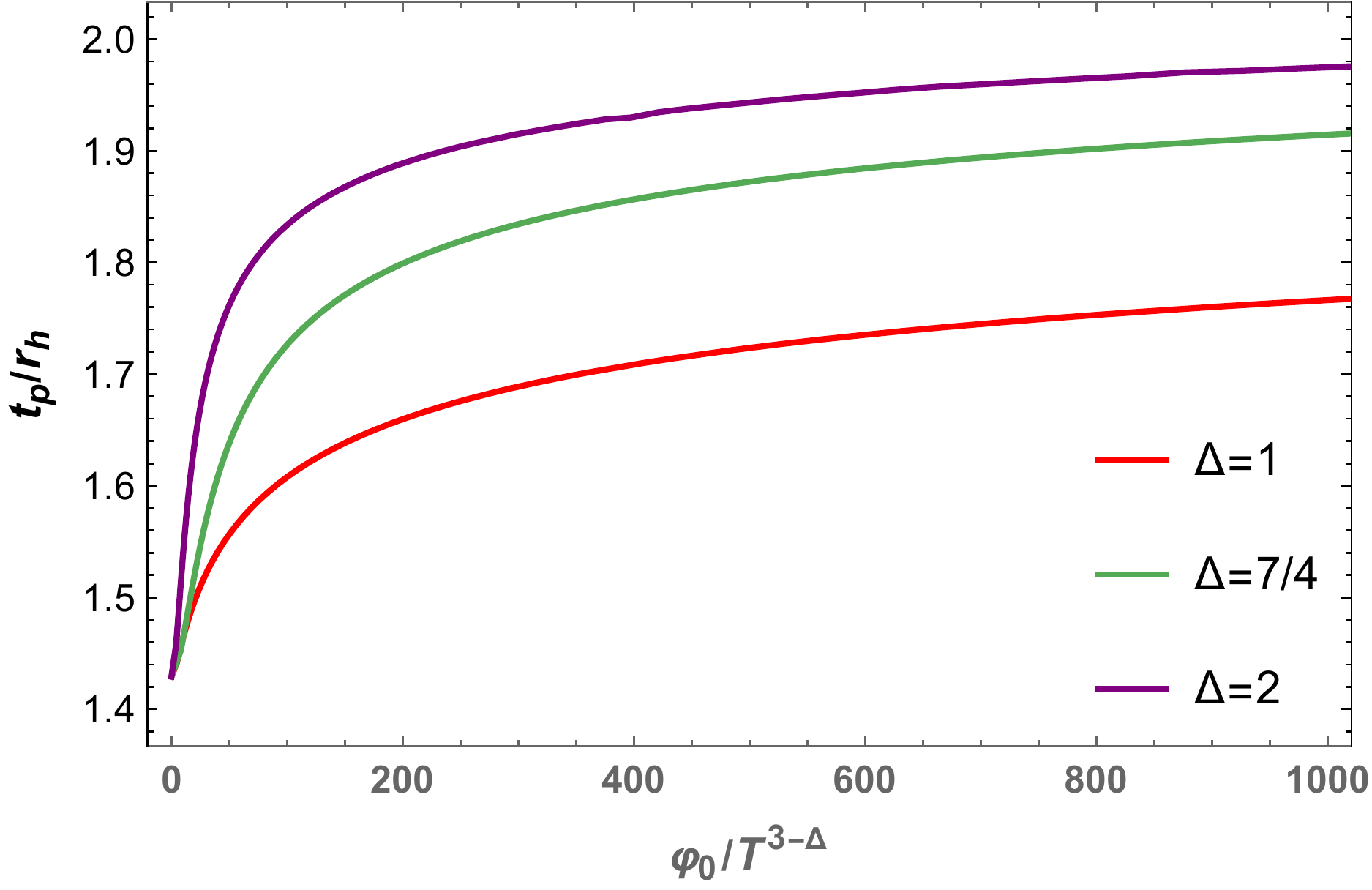}
			\caption{Page time with $\varphi_0/T^{3-\Delta}$.}
			\label{fig:fr}
		\end{subfigure}
		\caption{(a) RG flow from AdS boundary ($r \rightarrow 0)$ to the Kasner universe towards the singularity ($r \rightarrow \infty)$. Here $X$ denotes $\varphi(r)$, $\chi(r)$, $\ln g'_{tt}(r)$ and $f(r)$. The horizon is located as $r \approx 3.5$. (b) Variation of Page time with perturbation in $d=3$ for various scaling dimensions.}
		\label{fig:capae}
	\end{figure}

We also set $\chi_{h} = 0$. We now specialize for $d=3$. We solve Eqn.\eqref{eq1}-\eqref{eq3} and Eq.\eqref{ex1}-\eqref{ex3} with the conditions at horizon are given by Eq.\eqref{s1}-\eqref{s3}. For details, see \cite{Caceres:2021fuw}. One numerically obtains the solution of $f(r)$. Note that, $f(0)=1$, and the $f(r \approx 3.5) = 0$, which is the location of the horizon. We show the flow of $\varphi(r)$, $\chi(r)$, $\ln g'_{tt}(r)$ and $f(r)$ from AdS boundary to the Kasner universe towards the singularity in Fig.\ref{fig:fl}. Note that, towards Kasner singularity, the variation of the fields becomes constant, which gets fixed by the Kasner exponents.

\section{Bath deformations and Page time} \label{area}

This section briefly reviews the main results found in \cite{Caceres:2021fuw}. We take AdS$_{d}$ black hole geometry and couple it to a thermal bath. The fields at the interface obey transparent boundary conditions such that the black hole remains in thermal equilibrium with the bath all the time. We use prescription $(\text{III})$ to compute the entanglement entropy of the boundary subregion $\mathcal{R}$. In the next subsections, we elucidate the HM and island surface area computations, thereby obtaining the Page time as a function of the perturbation parameter.

\subsection{Area of the Hartman-Maldacena surface}
Before Page time, the HM surface is the dominant minimal surface. Due to the ever-growing nature of HM surface, its area i.e., the entanglement entropy also grows with time. To compute the entropy we take any $x_1=x_{\mathcal{R}}$ (constant) slice with the induced metric given by
\begin{equation}
    \mathrm{d}s^2|_{x_1=x_{\mathcal{R}}}=\frac{1}{r^2} \left[ - f(r) e^{-\chi(r)} \mathrm{d} t^2 + \frac{\mathrm{d} r^2}{f(r)}+\mathrm{d} \vec{x}^2_{d-2} \right].
\end{equation}
The area density functional for $r=r(t)$ is computed as
\begin{align}
     \mathcal{A} = \int\frac{\mathrm{d} t}{r(t)^{d-1}}\sqrt{-f(r(t))e^{-\chi(r(t))}+\frac{\dot{r}(t)^2}{f(r(t))}}=\int \mathrm{d}t \, \mathcal{L}.
     \label{HM area}
 \end{align}
where $\dot{r} = \mathrm{d}r/\mathrm{d}t$. Explicit time dependence is not present in $\eqref{HM area}$ so we can easily find the constant of the motion from the Lagrangian
 \begin{equation}
  \mathcal{E}=\dot{r}\frac{\partial \mathcal{L}}{\partial \dot{r}}-\mathcal{L}=\frac{f(r)e^{-\chi(r)}}{r^{d-1}\sqrt{-f(r)e^{-\chi(r)}+\frac{\dot{r}^2}{f(r)}}}.
 \end{equation}
This is noting but the energy of the corresponding minimal surface. The above equation can be rewritten in terms of the trajectory as
\begin{align}
    \dot{r} = \pm f(r) e^{-\chi (r)/2} \sqrt{1+ \frac{f(r) e^{-\chi(r)}}{r^{2(d-1)} \mathcal{E}^2}}. \label{traj}
\end{align}
Using this one computes the area of the HM surface given by
\begin{align}
    \mathcal{A}_{\mathrm{HM}}(t_b) = 2 \int_0^{\bar{r}} \frac{\mathrm{d} r}{r^{d-1} \sqrt{f(r) + e^{\chi(r)} r^{2(d-1)} \mathcal{E}^2}}, \label{hmarea0}
\end{align}
where $t_b$ is the boundary time and $\Bar{r}$ is such that $\dot{r}|_{r=\Bar{r}}=0$. This means at this point (from Eq.\eqref{traj}), we have $f(\Bar{r}) e^{-\chi(\Bar{r})} = - \Bar{r}^{2(d-1)}\mathcal{E}^2$. Further, the boundary time defined as
\begin{align}
    t_b = - P \int_0^{\Bar{r}} \mathrm{d} r \frac{ \mathrm{sgn}(E) e^{\chi(r)}/2}{r^{d-1} \sqrt{f(r) e^{-\chi(r)}/ (r^{2(d-1)} \mathcal{E}^2)}}.
\end{align}
Using this and Eq.\eqref{hmarea0} one can numerically compute the area of HM surface.

\subsection{Area of the island surface}

We find the anchoring surface (which is the island surface after Page time) in similar spirit to the calculation done in \cite{Bhattacharya:2021jrn}. We consider the $t=0$ slice given by
\begin{align}
    \mathrm{d} s^2|_{t=0} = \frac{1}{r^2} \left[ \frac{\mathrm{d} r^2}{f(r)} + \mathrm{d} \Vec{x}_{d-1}^2 \right].
\end{align}

The area functional we want to minimize is 
\begin{align}
    A = \int \mathrm{d} r \, \frac{1}{r^{d-1}} \sqrt{\frac{1}{f(r)} + x_1'(r)^2} = \int \mathrm{d} r \, \mathcal{L}, \label{lag1}
\end{align}
where $x_1' = \mathrm{d} x_1/\mathrm{d} r$ and $\mathcal{L}$ is the Lagrangian (density). It should be noted that we are considering area density as we have suppressed other $(d-2)$ transverse directions, which amounts to divide the area functional by $(d-2)$-dimensional volume. With the Lagrangian in Eq.\eqref{lag1}, we can find the equation of motion which has to be solved using the following boundary condition
\begin{align}
    x_1(0) = x_{\mathcal{R}}, ~~~~~ \frac{1}{x_1'(r)}\bigg|_{x_1 = 0} = 0.
\end{align}
Using this, we should be able to find the anchoring surface. However, as argued in \cite{Geng:2020fxl} and later in \cite{Ghosh:2021axl}, that for the fixed non-gravitating radiation region one can use the Dirichlet conditions and both conditions are equivalent. Hence, we extremize the action \eqref{lag1} with the following Dirichlet boundary condition
\begin{align}
    x_1 (r_T) = 0, ~~~~  x_1 (0) = x_{\mathcal{R}}, \label{bca}
\end{align}
where $x_R$ is the point located on the boundary and given by
\begin{align}
    x_{\mathcal{R}} = \int_0^{r_T} \frac{r^{d-1}\, \mathrm{d} r}{\sqrt{f(r)(r_T^{2(d-1)}-r^{2(d-1)})}}.
\end{align}
This allows us to solve the anchoring surface $x_{\text{Is}}(r)$
\begin{equation}
x_{\mathrm{Is}}(r)=x_{\mathcal{R}}-\int_{\epsilon}^{r}\frac{r^{d-1}\, \mathrm{d} r}{\sqrt{f(r)(r_T^{2(d-1)}-r^{2(d-1)})}}.
\end{equation}

In this way, one encounters two possible extremal surfaces in AdS$_{d+1}$ from which we choose the minimal one to get the unitary Page curve. This geometry is dual to the zero deformation field theory living on the boundary of the bulk geometry. The deformation of the bath state with a relevant scalar operator introduces a bulk scalar perturbation that deforms the near singularity regime to a more general Kasner universe. For any ${\varphi_0}/{T^{3-\Delta}}=\text{constant}$, we study the Page curves and find a unique Page time. Then by changing the deformation, we observe that the Page time becomes a monotonic function of ${\varphi_0}/{T^{3-\Delta}}$  shown in Fig.\ref{fig:fr}. The upshot is that the higher Page time results from an increased coarse-graining of the bath degrees of freedom. This makes the authors in \cite{Caceres:2021fuw} realize Page curves to probe the holographic RG flows.

\section{Bath deformations and subregion complexity} \label{complexitybath}

\begin{figure}
		\centering
		\begin{subfigure}[b]{0.48\textwidth}
			\centering
			\includegraphics[width=\textwidth]{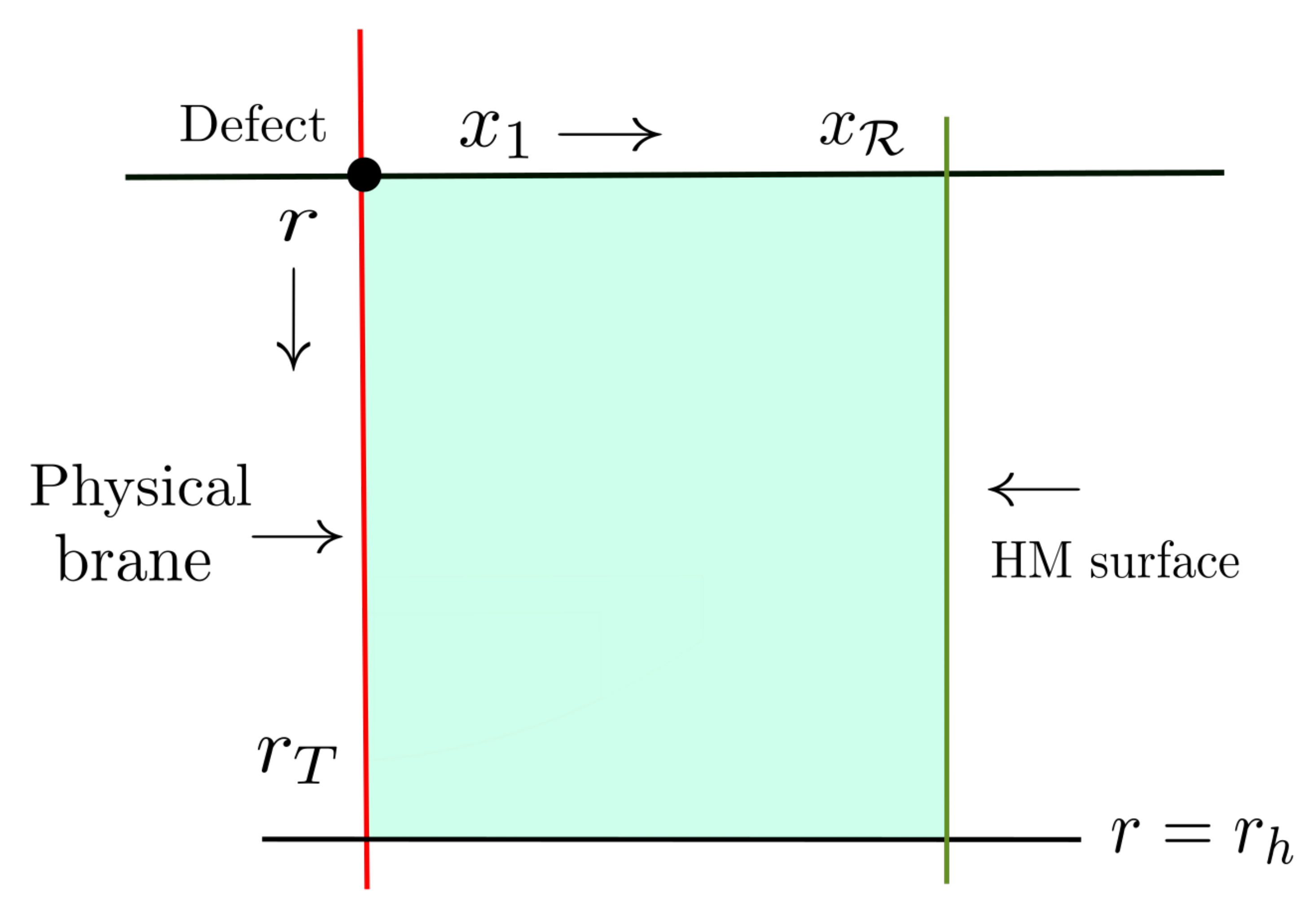}
			\caption{Volume under HM surface.}
			\label{fig:hm}
		\end{subfigure}
		\hfill
		\begin{subfigure}[b]{0.48\textwidth}
			\centering
			\includegraphics[width=\textwidth]{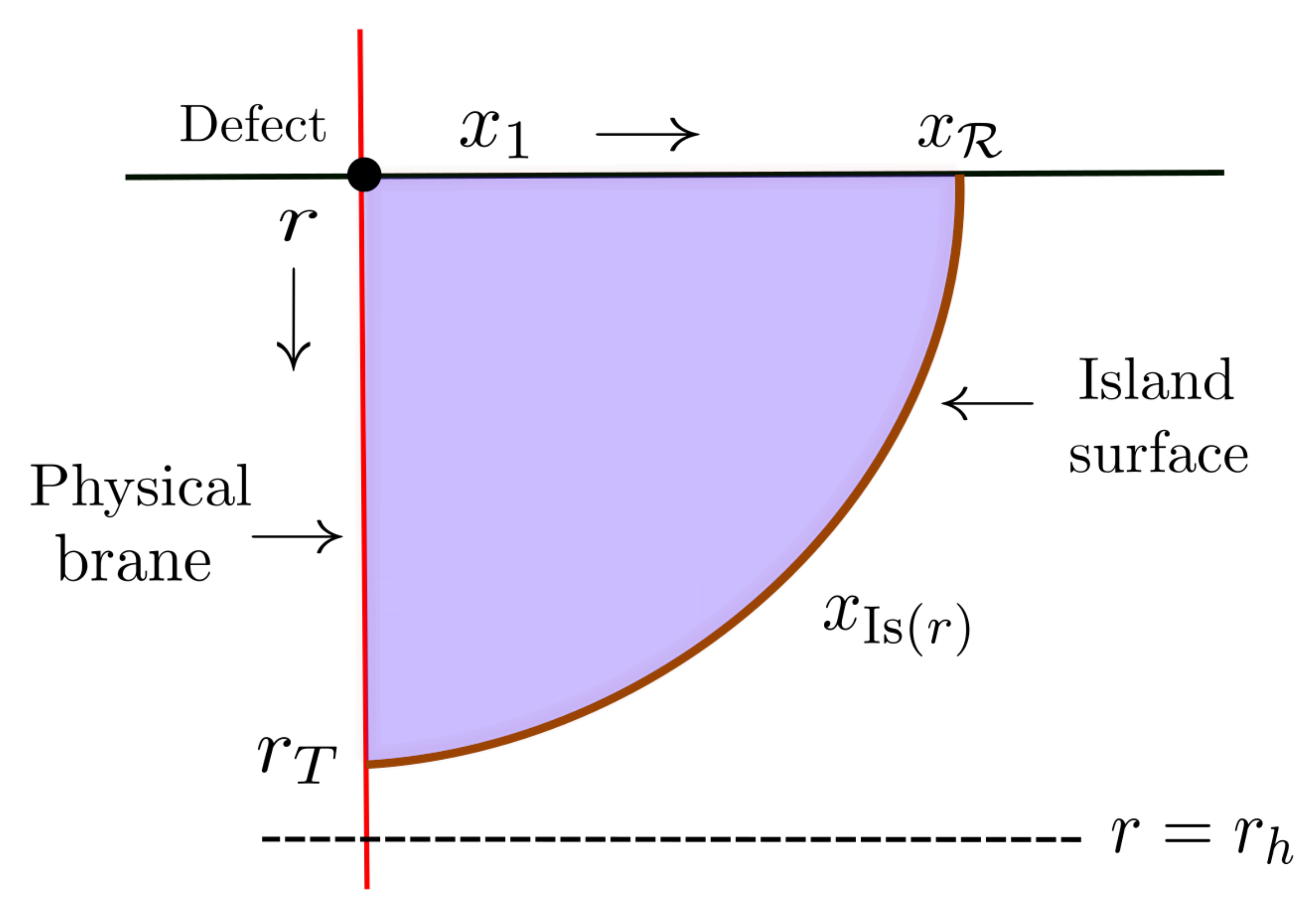}
			\caption{Volume under island surface.}
			\label{fig:is}
		\end{subfigure}
		\caption{(a) Volume under Hartman-Maldacena (HM) surface and the brane, marked by the green region.  (b) The island surface is denoted by $x_{\mathrm{Is}(r)}$. The purple region gives the volume under the island surface.}
		\label{fig:hmis}
	\end{figure}

As discussed before, the Hartman-Maldacena (HM) surface is the preferred RT surface before the Page time. Hence, the corresponding subregion complexity amounts to evaluate the volume between the HM surface and the brane. Refer Fig.\ref{fig:hm}, where the shaded green region shows the volume. However, at Page time, a transition happens. The constant island surface becomes the preferred RT surface. Hence, one computes the corresponding volume between the island surface and the brane as shown in Fig.\ref{fig:is} marked by the purple region. This section evaluates the corresponding volumes and subregion complexities before and after Page time for various deformations. As the transition of minimal surfaces happens at Page time, we expect a discontinuous behavior in complexity right at the Page time. We aim to observe this discontinuous jump for different perturbations to understand its nature better. However, before jumping to the main calculation, we briefly review the notion of holographic subregion complexity that we will be using in our following computations.

\subsection{Holographic subregion complexity}

The very idea of holographic subregion complexity is inspired by the proposal made in \cite{Alishahiha:2015rta}.\footnote{See \cite{Ben-Ami:2016qex, Carmi:2016wjl, Carmi:2017ezk, Abt:2017pmf, Abt:2018ywl, Chen:2018mcc, Du:2018uua, Auzzi:2019fnp, Auzzi:2019mah, Baiguera:2021cba, Auzzi:2021nrj, Bernamonti:2021jyu, Chapman:2021eyy, Emparan:2021hyr} for various computations of holographic and subregion complexity.} One starts by considering a static time slice and a subregion in the boundary and the corresponding minimal surface anchored in that boundary subregion. The area of the minimal surface is, of course, the entanglement entropy that we considered before. However, as soon as the minimal surface gets fixed, one can equivalently compute the volume enclosed by the minimal surface. This gives an alternate information-theoretic measure from the bulk point of view, which has been conjectured to be dual to the complexity of the boundary mixed states. The subregion complexity is defined in the purview of ``complexity=volume" proposal \cite{Susskind:2014moa, Susskind:2014rva, Carmi:2017jqz} as
\begin{align}
    C_{A} = \frac{V (A,\gamma_A)}{8 \pi \ell_{\mathrm{AdS}} G_N}. \label{cvsub}
\end{align}
where $V (A,\gamma_A)$ is the volume enclosed by the boundary subregion $A$, and the corresponding minimal surface is denoted by $\gamma_A$. Note that we are considering a static time slice and computing the volume enclosed by the minimal surface on this slice. For time-dependent cases, we need to resort to the covariant HRT proposal. However, here we focus only on the static case. Equipped with the above ideas, we are now able to calculate corresponding volumes and complexities where the minimal surfaces are the HM surface (before Page time) and island surface (after Page time).

\subsection{Volume under the Hartman-Maldacena surface}

\begin{figure}
    \centering
    \includegraphics[width=0.7\textwidth]{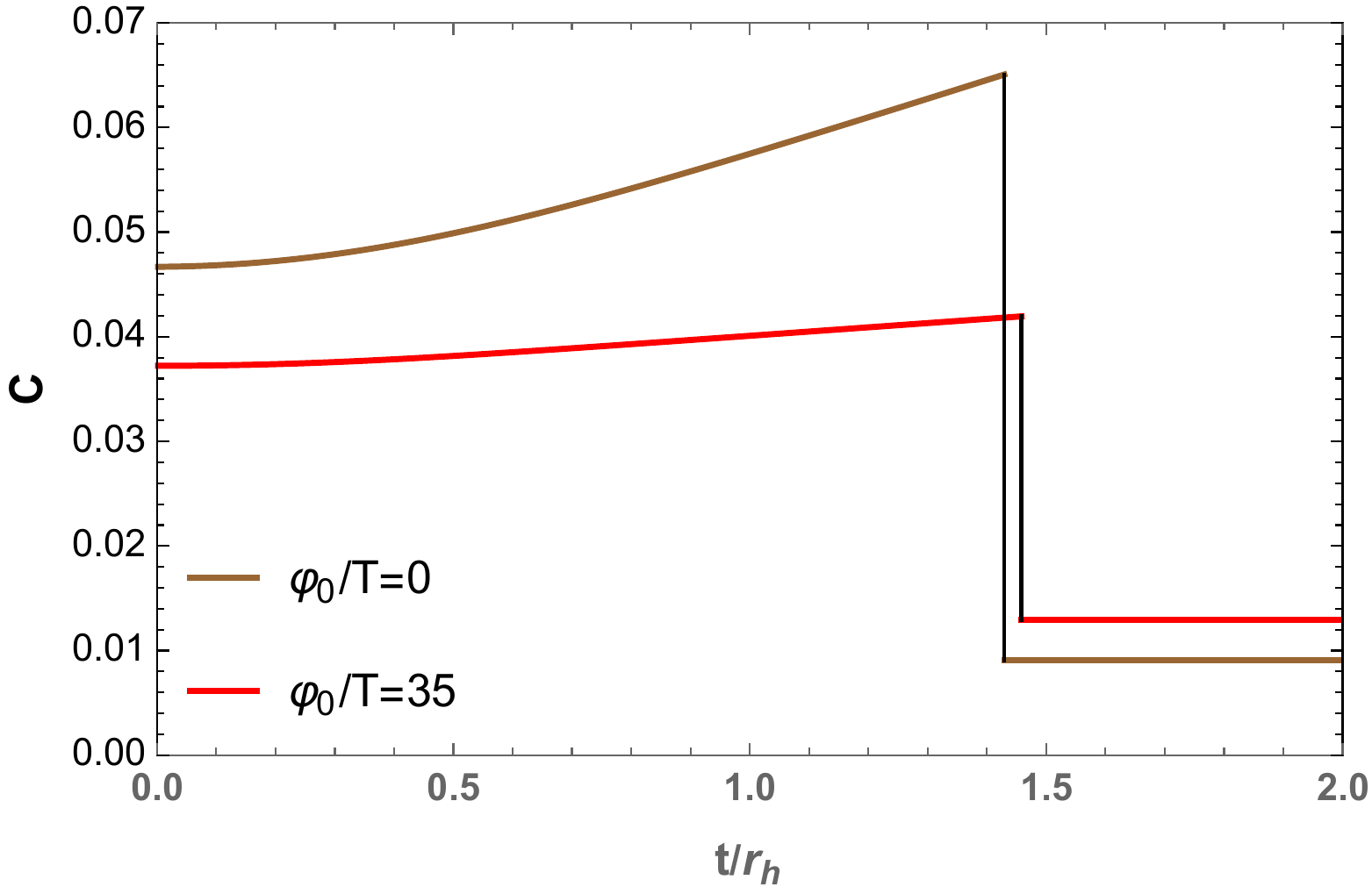}
    \caption{Evolution of complexity with time for $d=3$. The increasing nature of the complexity comes from the increasing growth of HM until Page time, after that complexity ceases to grow. This is due to the transition between two minimal surfaces at Page time. Therefore, complexity shows a discontinuous jump at Page time and captures the transition. The plot has been shown for two different perturbations, $\varphi_0/T = 0$ (zero-perturbation) and $\varphi_0/T =35$.}
    \label{comvst}
\end{figure}

We first compute the volume between HM surface and the brane. The HM surface and the island surface meet at the boundary interface point $x_\mathcal{R}$. We can foliate the region between the brane and the HM surface by infinite number of HM surfaces. We justify this by considering all possible $x_1=\text{constant}$ slices
\begin{equation}
   \mathrm{d} s^2|_{x^1=\text{constant}}=\frac{1}{r^2} \left[ - f(r) e^{-\chi(r)} \mathrm{d} t^2 + \frac{\mathrm{d} r^2}{f(r)}+\mathrm{d} \vec{x}^2_{d-2} \right].
\end{equation}
 The minimal surface for each such $x_1=\text{constant}$ slice is a HM surface. Therefore, we can choose it to foliate the region between the brane and the HM surface located at $x_1=x_\mathcal{R}$. The ever-growing area of the HM surface is given by
 \begin{align}
     \mathcal{A} = \int\frac{\mathrm{d} t}{r(t)^{d-1}}\sqrt{-f(r(t))e^{-\chi(r(t))}+\frac{\dot{r}(t)^2}{f(r(t))}},
 \end{align}
 which has been computed in Eq.\eqref{HM area}. However, here we are interested in calculating the subregion volume density. This is given by
\begin{equation}
\mathcal{V}_{\mathrm{HM-Br}}(t_b)=2\int\frac{\mathrm{d} t\,\mathrm{d} x_1}{r(t)^d}\sqrt{-f(r(t))e^{-\chi(r(t)}+\frac{\dot{r}(t)^2}{f(r(t))}}=2x_\mathcal{R}\int\frac{\mathrm{d} t}{r(t)^d}\sqrt{-f(r(t))e^{-\chi(r(t))}+\frac{\dot{r}(t)^2}{f(r(t))}}.
\label{VHM}
\end{equation}
We substitute the HM solution in $\eqref{VHM}$ and find the volume density,
\begin{equation}
  \mathcal{V}_{\mathrm{HM-Br}}(t_b)=2x_\mathcal{R}\int_{\epsilon}^{\bar{r}}\frac{1}{r^d}\frac{\mathrm{d} r}{\sqrt{f(r)+e^{\chi(r)}(r^{d-1}E)^2}}.
\label{VHM1} 
\end{equation}
We numerically evaluate the volume and the corresponding subregion complexity using Eq.\eqref{cvsub} and shown in Fig.\ref{comvst}. The HM surface is ever-increasing. Hence the volume enclosed by the HM surface and the brane will also be increasing with time. This increasing nature persists up to the Page time and can be seen from Fig.\ref{comvst}. Here we have computed the evolution for two different perturbations namely $\varphi_0/T = 0$ (zero-perturbation) and $\varphi_0/T =35$ at $d=3$. We see that, with the increasing growth of perturbation, the growth of volume decreases. Finally, the HM surface ceases to become the minimal surface at Page time, and the island surface takes over. However, the HM still grows with time.

\subsection{Volume under the island surface}\label{ivolume}
In the previous section, we have computed the volume between the Hartman-Maldacena (HM) surface and the brane.  This subsection evaluates the volume between the island surface and the brane. As discussed before, the constant island surfaces are the preferred RT surfaces after the Page time.

The subregion complexity from the black hole perspective is given by the volume enclosed by the island surface and the physical brane. At $t=0$, $\eqref{met}$ becomes
\begin{equation}
    \mathrm{d} s^2|_{t=0} = \frac{1}{r^2} \left[ \frac{\mathrm{d} r^2}{f(r)} + \mathrm{d} \Vec{x}_{d-1}^2 \right].
\end{equation}
The volume density under the RT surface which is depicted in the Fig.\ref{fig:is} is
\begin{equation}
\mathcal{V}_{\mathrm{Is-Br}}=2\int\frac{1}{r^d}\frac{\mathrm{d} r\, \mathrm{d} x_{1}}{\sqrt{f(r)}}=2\int_{\epsilon}^{r_T}\frac{x_{\mathrm{Is}}(r)}{r^d \sqrt{f(r)}}\, \mathrm{d} r.
\label{Isbr}
\end{equation}
where the embedding function $x_{\mathrm{Is}}(r)$ is,
\begin{equation}
x_{\mathrm{Is}}(r)=x_{\mathcal{R}}-\int_{\epsilon}^{r}\frac{r^{d-1}\, \mathrm{d} r}{\sqrt{f(r)(r_T^{2(d-1)}-r^{2(d-1)})}}.
\end{equation}
We again numerically calculate this volume, and the result is shown in Fig.\ref{comvst}. However, there is a difference from the previous results obtained for the HM surface. The volume enclosed by the island surface and the brane is constant in time and dominates after the Page time. See Fig.\ref{comvst}. Here, we have similarly computed the complexities for two different perturbations, $\varphi_0/T = 0$ (zero-perturbation) and $\varphi_0/T =35$ at $d=3$.

The overall evolution of complexity shows the following pattern. When the ever-growing HM surface is minimal, complexity grows over time. At Page time, island surfaces become the preferred minimal surface and continue to be. Hence, the transition between the two minimal surfaces is well computed by the discontinuous jump of the corresponding subregion complexities.

	\begin{figure}
		\centering
		\begin{subfigure}[b]{0.49\textwidth}
			\centering
			\includegraphics[width=\textwidth]{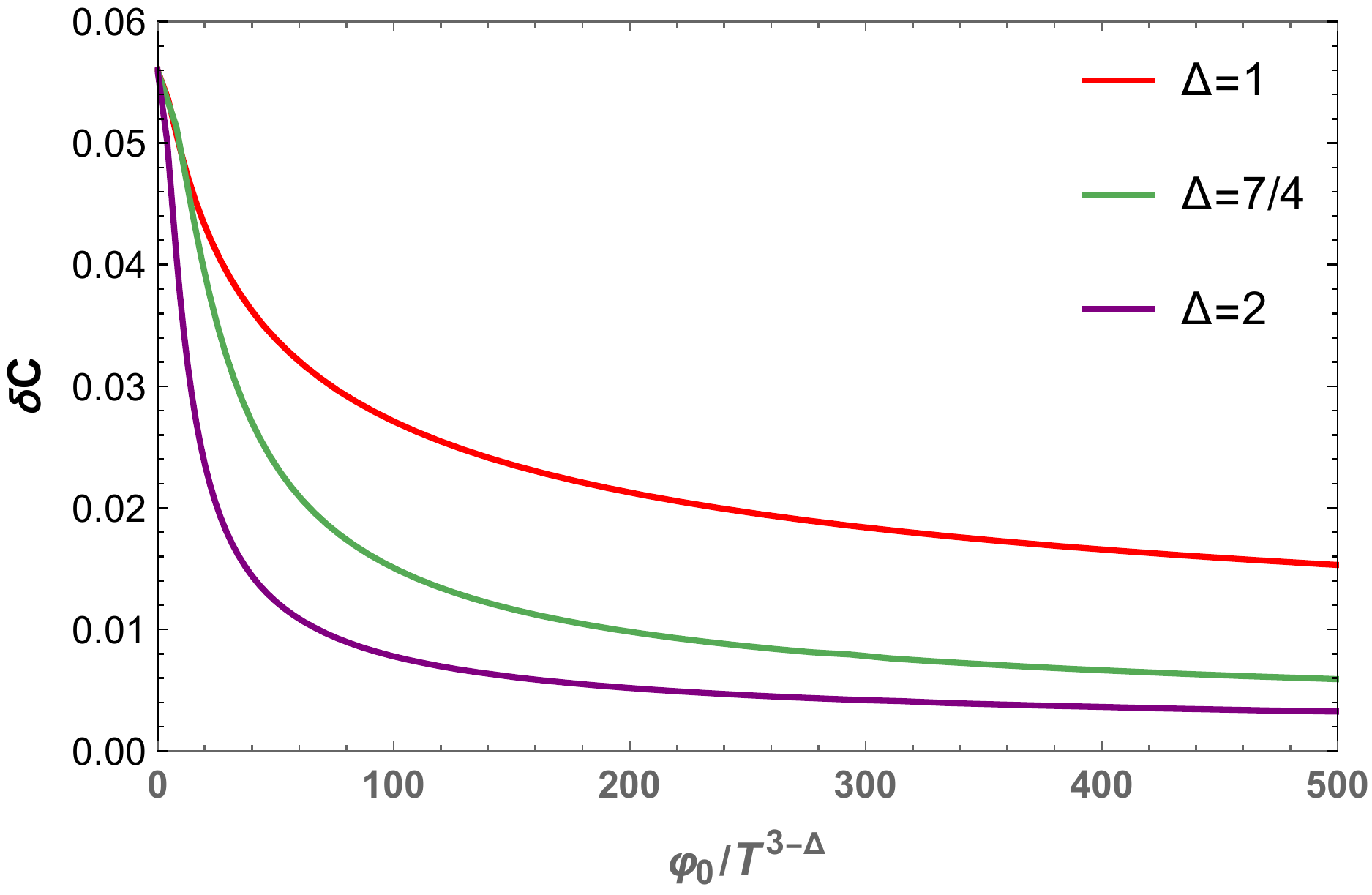}
			\caption{$\delta C$ with $\varphi_0/T^{3-\Delta}$.}
			\label{fig:cc1}
		\end{subfigure}
		\hfill
		\begin{subfigure}[b]{0.49\textwidth}
			\centering
			\includegraphics[width=\textwidth]{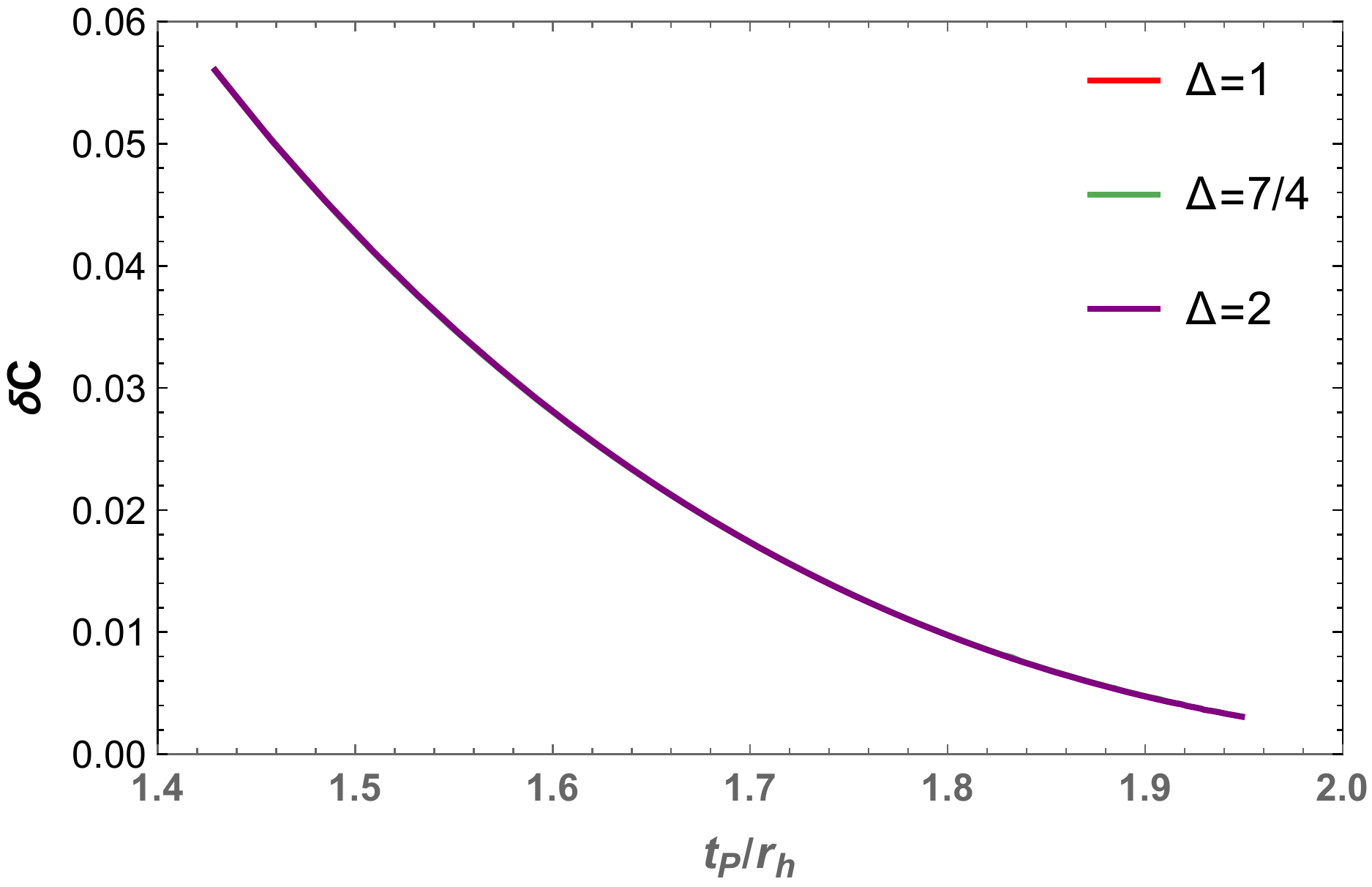}
			\caption{$\delta C$ with $t_P/r_h$.}
			\label{fig:cc2}
		\end{subfigure}
		\caption{(a) The discontinuous behavior of complexity with the perturbation for different scaling dimensions. (b) The jump in complexity with Page time. For all scaling dimensions, the pattern decreases and superimposes on each other. More explanation is given in the text.}
		\label{fig:cc}
	\end{figure}

To understand this discontinuous jump quantitatively, we define the difference between $\eqref{VHM1}$ and $\eqref{Isbr}$, which is a UV finite quantity, $\delta V$ at Page time as
\begin{equation}
  \delta \mathcal{V}=\mathcal{V}_{\mathrm{HM-Br}}-\mathcal{V}_{\mathrm{Is-Br}}
  \label{VDiff}
\end{equation}
We again stress that we are interested in computing $\eqref{VDiff}$ at $t=t_P$ (page time) because there is a discontinuity in the holographic subregion complexity at the Page time due to the auto-purification. To understand its nature, an obvious way to deform the bath state by relevant perturbations and realize the jump for various perturbations. Specifically, we study the jump in complexity at $d=3$, for different scaling dimensions. The result is shown in Fig.\ref{fig:cc1}. We see that, for a fixed scaling dimension, the discontinuity decreases with the perturbation, consistent with Fig.\ref{comvst}. Further, as we increase the scaling dimensions, the corresponding jump in complexity decreases for a fixed perturbation.

We also consider the discontinuous nature of complexity with respect to Page time, shown in Fig.\ref{fig:cc2}. We see that the discontinuity decreases with Page time. However, we get the same plot of all $\Delta$, implying that the decreasing behavior is independent of scaling dimensions. This is plausible, as the Page time also depends on the perturbation \cite{Caceres:2021fuw}.


\begin{figure}
    \centering
    \includegraphics[width=0.82\textwidth]{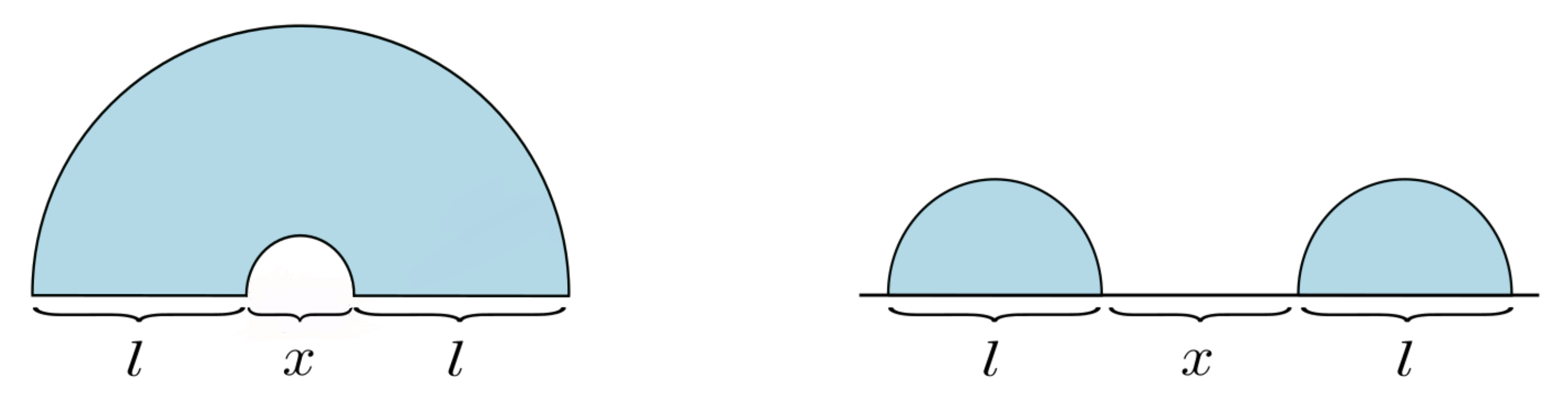}
    \caption{Two strips of length $l$ are separated by distance $x$, in general, $d$-dimensions. In this picture, the strips are drawn by lines. As we change the separation between them, we observe a transition between the minimal surfaces, and as a result, the enclosed volume jumps discontinuously.}
    \label{carmi}
\end{figure}

\begin{figure}
    \centering
    \includegraphics[width=0.68\textwidth]{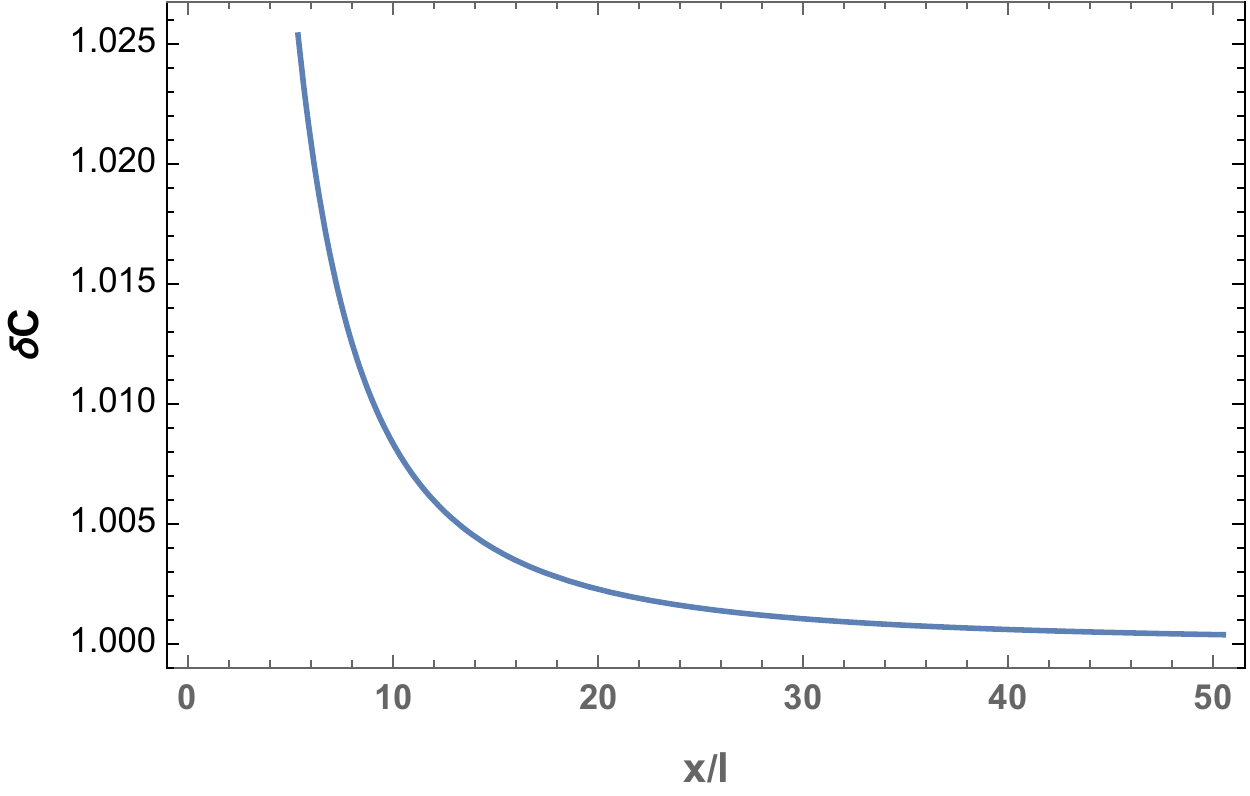}
    \caption{The difference $\delta C$ with respect to $x/l$ for AdS$_4$ $(d=3)$. Here we have chosen $l=2$. The plot has a close similarity with Fig.\ref{fig:cc1}.}
   \label{trans}
\end{figure}

It is interesting to see how the discontinuous behaviors of volumes (and hence, subregion complexities) arise in a simple setup. To see this, consider two strips of length $l$, separated by a distance $x$ as shown in Fig.\ref{carmi} in $d$-dimensions. When the strips are close (right), the entanglement wedge is connected, and the volume is also connected. However, when they are far apart (left), the corresponding complexities are given by the volumes of two disconnected pieces.  Hence, as we increase the separation, we transit from connected to the disconnected phases, and there is a discontinuous jump in the corresponding volume. However, here we do not consider any transition; rather, we think the connected and disconnected geometries exist on their own. This is reasonable, as, for any value of $x$, we can always consider a connected geometry, even if it is not a minimal surface. We are only interested in computing the volumes of connected and disconnected geometry for any separation $x$ and try to see how the difference of their respective volumes behaves as we increase the separation. We follow \cite{Ben-Ami:2016qex} to compute these volumes. The connected geometry has the volume
\begin{align}
    \mathcal{V}_{\mathrm{c}} = - \frac{c_0}{(x + 2l)^{d-1}} + \frac{c_0}{x^{d-1}},
\end{align}
where $c_0$ is a constant \cite{Ben-Ami:2016qex}. However, the disconnected geometry consists of two identical pieces, which together has the volume
\begin{align}
     \mathcal{V}_{\mathrm{dc}} = - \frac{2c_0}{l^{d-1}}.
\end{align}
Hence, the volume difference, and hence the difference in complexity in connected and disconnected geometry can be computed as \cite{Ben-Ami:2016qex}
\begin{align}
    \delta C = C_{\mathrm{c}} - C_{\mathrm{dc}} = c_0 \bigg[ - \frac{1}{(x + 2l)^{d-2}} + \frac{1}{x^{d-2}} + \frac{2}{l^{d-2}} \bigg]
\end{align}
This volume (complexity) is always positive, as the enclosed volume of the connected region is always greater than that of the two separate regions. We again stress that we do not consider any transition from connected to disconnected geometry, so we do not talk about any entanglement wedge. We can see how this difference $\delta C$ behaves as we increase the separation $x$ for a fixed $l$. In Fig.\ref{trans}, we have plotted $\delta C$ with respect to $x/l$ for AdS$_4$.

 The close similarity between Fig.\ref{fig:cc1} and Fig.\ref{trans} is striking. Note that the asymptotic value in Eq.\eqref{trans} depends on the value of $l$, which is constant.  The discontinuity of volume behaves similarly, if one considers $x/l$ as analogous to the perturbation $\varphi_0/T$.\footnote{One can think that $x$ might play the role of time here, but as we have argued before, we are not considering any transition. We are simply sitting at the Page time. Hence, $x$ does not play the role of time.}  One way to think about this particular behavioral similarity is by considering the size of the radiation subsystem. This can be understood if we consider another set of RG flow by changing the size of the radiation subsystem, which is supposed to be another valid parameter that can change the Page time and hence the jump of complexity at Page time as well. In that case, from this simple example of disjoint subsystem volumes, we can intuitively infer a similar relation between the Page time and the subsystem size ($x/l$ in this case is a little different as in this particular example, the distance between two disjoint subsystems plays the role of the parameter which changes. In case of subsystem size, $l$ will be the parameter one is supposed to vary). Another interesting choice could be taking disjoint subsystems in the radiation side and increase/decrease both of them along some RG flow. Overall the expectation would be that in such a RG flow as well, the transition between two competing minimal surfaces will take place at Page time. Page time will have similar growth and saturation behavior as shown in Fig.\ref{fig:fr}, but with some parameter similar to $x/l$. This also intuitively indicates that for different sets of variables that can induce a RG flow, the behavior of the jump of volumes show some kind of universal behavior. However, it should be noted that, in the braneworld model, there is no disconnected regions in the bath. Hence, one should not think a one to map between two different descriptions. They are described purely as an analogy, especially how to think the discontinuous behavior of complexity in the braneworld model in a simple geometric way.

\section{Discussions and outlook} \label{discussion}

In this paper, we have studied the holographic subregion complexity in a braneworld model with various deformation in the bath degrees of freedom. More specifically, we deform the bath state by a relevant scalar operator. This, in turn, induces a renormalization group flow in the bulk state from AdS-Schwarzchild to Kasner universe. We study the corresponding model in a doubly holographic setup. For each relevant perturbation, the model consists of two competing minimal surfaces: the Hartman-Maldacena (HM) surface and the island surface. We investigate the behavior for a class of such RG flows induced by the parameters of the relevant deformations and check how the deformation parameters change the volumes under the HM and island surfaces. It was previously found that as the parameter under study (in our case, this is $\varphi_0/T^{d-\Delta}$) is increased, the Page time (where the quantum extremal surface goes through a transition) increases initially before achieving a saturation. We have studied the dependence on the dimension of the operator inserted in the boundary ($\Delta$) with dimension $d=3$. The growth of the Page time is increased as we increase the operator dimension. It also saturates to a higher value in such a case. These behaviors regarding Page time were found primarily in \cite{Caceres:2021fuw}. Our main focus was to study the jump (positive from radiation side and negative from black hole side) in subregion volume at Page time along with these deformations. Our main results are summarised below.

\begin{enumerate}
    \item The Hartman-Maldacena volume associated with the black hole side decreases with increasing $\varphi_0/T$ whereas the island volume increases for the same case (as shown in Fig.\ref{comvst} with $d-\Delta=1$). This clearly indicates the fact that the dip at Page time decreases with increasing deformation.
    
    \item When we study the behavior of the jump $\delta C$ extensively with varying deformations, we find that it decreases initially with increasing deformation and saturates at a later point, similar to the saturation of Page time. The rate of decrease increases with the increasing value of the operator dimension. Again this is similar in spirit to the behavior of Page time, in which case the rate of increase increases with increasing operator dimension.
    
    \item Motivated by these similarities, when we plot $\delta C$ vs. $t_{P}/r_h$, we find that increasing Page time indeed seems to decrease the jump at Page time. However, the Page time and the jump both are present even in the absence of such a deformation picture. This motivates us to conjecture that the jump should depend on the Page time in a similar model even in the absence of such deformation parameters. This understanding is crucial because in all the previous studies concerning this jump (\cite{Bhattacharya:2020uun, Bhattacharya:2021dnd, Bhattacharya:2021jrn}), the Page time and the jump was studied numerically, and there was no clear way to understand this relation between them. We believe this result is a model-independent fact.
    
    \item Another interesting and important observation is that the $\delta C$ vs. $t_{P}/r_h$ plots for different $\Delta$ overlap with each other. This also supports our conjecture about the universality of the relation between the jump and the Page time. It reflects the fact that even if we change the operator dimension through which the deformation is introduced, the slope of the $\delta C$ vs. $t_{P}/r_h$ curve does not change. This universality also relates these different sets of RG flows in a way. It means that this particular ratio is invariant of the operator dimension. It will be interesting to understand the implication of this fact in more detail.
    
\end{enumerate}

In section \ref{ivolume}, we have also tried to show an analogous behavior of complexity jump in a simple setup where the distance between two joint subsystems in the boundary is varied. The nature of the plot is closely similar to what we find in the case of the complexity jump versus deformation parameter plot. We use this example behavior to argue that this indicates that other parameters like the size of the radiation subsystem could also change this jump at Page time. In this paper, we observed the change of the jump along with a set of RG flows that changes the Page time, and hence we can comment on how the jump might depend upon the Page time. It would be interesting to model a different set of RG flows by changing the size of the radiation subsystem and check if a similar behavior persists between the jump of complexity at Page time and the subsystem size. One could also consider disjoint subsystems and vary both the size of the subsystems and the distance between them. In general, it remains an open problem to fully understand how the Page time and hence the jump of complexity at that time depend on different parameters of the theory. 

In \cite{Bhattacharya:2020uun}, authors studied complexity of the radiation subsystem and in that case the jump at Page time is positive. This jump is exactly same in magnitude to the negative jump in case of black hole subsystem as was found in \cite{Bhattacharya:2021jrn}. The authors of \cite{Bhattacharya:2020uun} attributed this jump in complexity of radiation to the purification of modes between island region and radiation region. The idea is that although field theoretically the mixed state complexity we are studying is supposed to be given by complexity of purification, there is an extra auto-purification of certain modes going on at Page time due to inclusion of the islands. For purification of the other modes, one still has to add auxiliary system in a field theory setup whereas the island modes act as purifying partners of certain modes for which one does not need to add any auxiliary system by hand. This auto-purification results in the extra complication (simplicity) at Page time for the radiation (black hole) subsystem. Let us assume that this extra set of gates that one needs to introduce to mimic the auto-purification, is the equivalent of the magnitude of this jump from a field theoretic point of view. Our results from this paper suggests that the number of these auto-purifying gates decreases as the relevant deformation is increased. We believe that this means the number of modes getting purified due to the inclusion of islands also decreases with increased deformation. This can be roughly thought of as the island region containing less and less number of radiation partner modes in them as the deformation is increased. However this argument is a little bit stretched. One would need a far better understanding of what is happening physically in these island models from a field theoretic point of view to check this argument from explicit field theoretic computations.

An interesting direction to extend our study would be to examine the covariant proposal directly instead of taking a fixed time slice. This case has subtle issues that were previously addressed in \cite{Bhattacharya:2021jrn}. An outstanding question will be to understand the corresponding jump in subregion complexity and its dependence on the perturbation parameters if one adds a second brane in the model \cite{Geng:2020fxl} and considering a gravitating bath. Further, recently a formulation of the subregion volume has been given in terms of bit threads \cite{Pedraza:2021mkh, Pedraza:2021fgp}. The exciting open question is to understand whether bit thread formulation can provide a more vivid picture of the discontinuity of the subregion complexity in the doubly holographic braneworld model, perhaps along the line of \cite{Rolph:2021hgz, Agon:2021tia}.

\section*{Acknowledgements}
We thank Arnab Kundu for useful discussions. A.B.$(1)$ is supported by the Institute of Eminence endowed postdoctoral fellowship offered by Indian Institute of Science. A.B.$(2)$ is supported by Start-Up Research Grant (SRG/2020/001380) by the Department of Science \& Technology Science and Engineering Research Board (India) and Relevant Research Project grant (58/14/12/2021-BRNS) by the Board Of Research In Nuclear Sciences (BRNS), Department of Atomic Energy, India. P.N. acknowledges the University Grants Commission (UGC), Government of India, for providing financial support. A.K.P. is supported by the Council of Scientific \& Industrial Research (CSIR) Fellowship No. $09/489(0108)/2017$-EMR-I.

\bibliographystyle{JHEP}
\bibliography{references}

\end{document}